\documentclass[superscriptaddress,floatfix,prl,twocolumn,amsmath,amssymb]{revtex4-1}

\usepackage{graphicx}
\usepackage{dcolumn}
\usepackage{bm}
\usepackage{amssymb}
\usepackage{natbib}
\usepackage{amsmath}
\usepackage{mathtools}
\usepackage{color}
\usepackage{datetime}
\usepackage{footnote}
\usepackage{bbold}
\usepackage[normalem]{ulem}
\usepackage{dcolumn}
 \usepackage{ragged2e}
 \usepackage{xfrac}
 \usepackage{sidecap}
 \usepackage{setspace}
 \usepackage{multirow}
 \usepackage{xcolor}
 \usepackage{flushend}
\usepackage{hyperref}
\usepackage{cleveref}
\usepackage{microtype}
\usepackage{siunitx}

\begin{document}

\title{Ultra-low loss photonic circuits in Lithium Niobate On Insulator}
\author{Inna Krasnokutska}
\thanks{These authors contributed equally to this work}
\affiliation{Quantum Photonics Laboratory and Centre for Quantum Computation and Communication Technology, School of Engineering, RMIT University, Melbourne, Victoria 3000, Australia}

\author{Jean-Luc J. Tambasco}
\thanks{These authors contributed equally to this work}
\affiliation{Quantum Photonics Laboratory and Centre for Quantum Computation and Communication Technology, School of Engineering, RMIT University, Melbourne, Victoria 3000, Australia}

\author{Xijun Li}
\affiliation{Quantum Photonics Laboratory and Centre for Quantum Computation and Communication Technology, School of Engineering, RMIT University, Melbourne, Victoria 3000, Australia}

\author{Alberto Peruzzo}
\affiliation{Quantum Photonics Laboratory and Centre for Quantum Computation and Communication Technology, School of Engineering, RMIT University, Melbourne, Victoria 3000, Australia}


\begin{abstract}
Lithium Niobate on insulator (LNOI) photonics promises to combine the excellent nonlinear properties of Lithium Niobate with the high complexity achievable by high contrast waveguides.
However, to date, fabrication challenges have resulted in high-loss and sidewall-angled waveguides, limiting its applicability.
We report LNOI single mode waveguides with ultra low propagation loss of \SI{0.4}{dB/cm} and sidewall angle of \SI{75}{\degree}.
Our results open the route to a highly efficient photonic platform with applications ranging from high-speed telecommunication to quantum technology. 
\end{abstract}

\maketitle

\section{Introduction}
Integrated photonics is widely employed in high bandwidth telecommunication \cite{Hu:2012}, frequency conversion and filtering \cite{Wang:2016gk, Geiss:2015ia}, biophotonics and sensing \cite{Halldorsson:10}, and single photon generation and manipulation for quantum technology \cite{O'Brien:2012, Alibart:2016jo, Zeilinger2002}.
Important requirements of an attractive photonic platform are low propagation loss, high nonlinearities, high index contrast and industry compatible fabrication processes. A major player is silicon (Si) photonics, enabling very compact, low loss waveguides that can be fabricated with CMOS technology. Si has no second order nonlinearities, limiting the performance of important photonic components such as optical switches and frequency converters \cite{Lipson:2005}, and it absorbs light below \SI{1}{\micro\metre} wavelength, precluding its application in biophotonics. This has lead to research into other materials including InP\cite{Stabile:2016}, SiN\cite{Bauters:11}, GaAs\cite{Fiore:2016} and AlN\cite{Tang:2012} as photonic platforms. InP, GaAs and SiN all represent promising solutions for scalable and low-loss photonics; however, their switching capabilities are also limited due to small or absent second order nonlinearities and rely upon thermally unstable or absorptive switching mechanisms \cite{Stabile:2016,Fiore:2016}. Although AlN enables the use of electro-optical properties and frequency conversion, the low second order nonlinearities still limits the efficiency of these processes\cite{Tang:2012}.  Low loss is another major requirement for scalability, especially in quntum photonics.  Photonic components, such as rings, are important for filtering and delay lines and require a high quality factor, and therefore a propagation loss is vital to their operation.

Lithium Niobate (LN) has several potential advantages over competing platforms including a broad transparency range from \SI{350}{\nano\metre} to \SI{5200}{\nano\metre}, with potential application from biophotonics to mid-IR; a high electro-optic coefficient, enabling efficient ultra-fast optical switches; piezoelectric and pyroelectric properties, it can be periodically poled for wavelength conversion and single photon generation \cite{Keller:2016, Alibart:2016jo}; and it can be doped with erbium atoms to create waveguide integrated lasers \cite{Sohler:2000}.
Waveguides in Lithium Niobate have been fabricated via titanium in-diffusion (Ti:LN)--the industry standard for photonic modulators \cite{Janner:2009}--and proton-exchanged (PE:LN) \cite{Micheli:2014}. Both Ti:LN and PE:LN suffer from low refractive index contrast waveguides, greatly limiting the complexity of the photonic circuitry on these platforms.

To achieve high index contrast waveguides, etching of PE:LN\cite{Hu:2006kn}, hybrid-integration with Si \cite{Rabiei:2013kv,Witmer:2016vz} and SiN \cite{7350128,Chang:17}, as well as blade dicing \cite{Volk:2016ga} and micromachining \cite{Takigawa:2014kl}, have been reported.  Recently, the ability to create high quality LN thin films on SiO$_2$ insulator (LNOI) via the smart-cut technique \cite{Gunter:2012}, has enabled the direct fabrication of waveguides using standard lithography and dry etching techniques, with reported propagation loss as low as \SI{3}{\deci\bel/\centi\metre}\cite{Wang:2016uy}. To the authors' knowledge, all high-index contrast LNOI waveguides reported to date exhibit either a high loss \cite{Hu:09}, a shallow sidewall angle \cite{Wang:2016uy,Hu:09,Wang:14}, limited etch depth \cite{Wang:14}, or a combination thereof. 
Propagation losses in LN are almost entirely dominated by the quality of the nanofabrication because LN has very low intrinsic absorption. 
Several techniques have been attempted to micro- and nano-process LN, including ion-beam enhanced etching \cite{Geiss:2015ia}, wet etching with hydrofluoric acid (HF) based etchants \cite{Hu:hz} and reactive ion etching (RIE). RIE possesses particularly anisotropic properties; however, plasma etching LNOI is challenging. LN is highly reactive with fluorine (F) gases making them a logical choice for achieving good etch rates, but unfortunately, LiF products of this reaction deteriorate the surface leading to high scattering loss. An alternative to chemical RIE is argon milling, but this process has poor etch-selectivity, making it difficult to find a suitable mask, and is well known to result in very shallow sidewalls. Near vertical sidewalls are critical to achieving low loss waveguides, as well as high free spectral range (FSR) rings and small-footprint optical components, such as switches and couplers.

Here we report LNOI waveguides with propagation loss $\sim$0.4 dB/cm, achieved by an optimized etching process that produces a sidewall roughness of $<$\SI{2}{nm} RMS and a sidewall angle of $\sim$\SI{75}{\degree}. Our low loss ridge waveguides are fabricated using standard nanoprocessing techniques and enable the development of high density nonlinear and electro-optic based photonic circuitry.

\section{Fabrication}

\begin{figure}[h]
\centering
\includegraphics[width=1.0\linewidth]{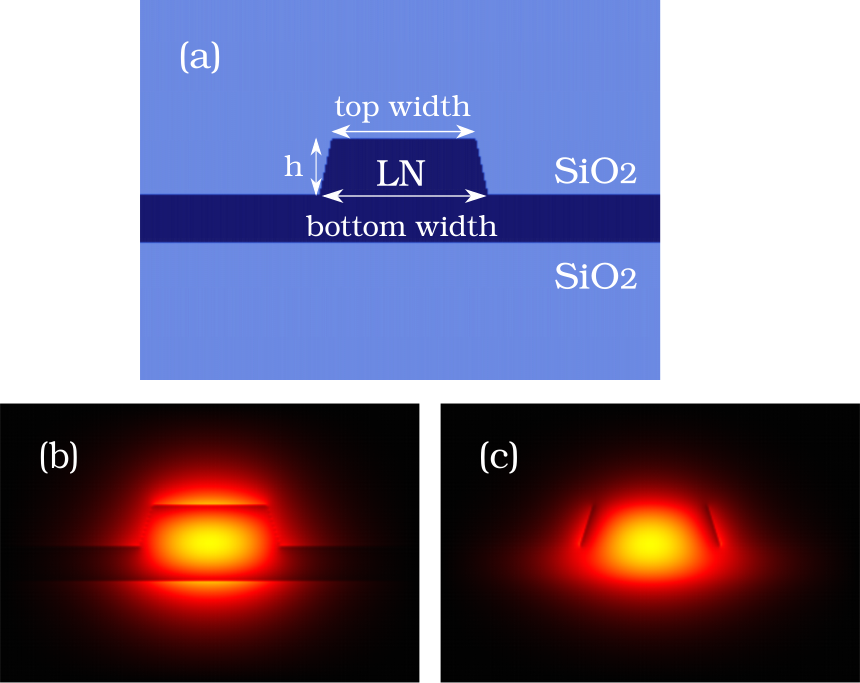}
\caption{Design of LNOI waveguides: (a) schematic cross-section of z-cut LNOI waveguide; (b) electromagnetic field distribution for TE mode; (c) electromagnetic field distribution for TM mode.}
\label{fig1}
\end{figure}

A mode solver was used to determine the dimensions of a \SI{1550}{\nano\meter} single mode ridge waveguide in LNOI. The design of the waveguide includes the following parameters: ridge height, bottom width, top width, refractive indices of the waveguide and claddings, and film thickness.  \Cref{fig1}(a) shows a typical cross-section of a z-cut ridge waveguide with SiO$_2$ cladding. 

The difference in the waveguide dimensions at the top and bottom of the ridge is due to the sidewall angle introduced by the etching process. LNOI waveguides support both TE and TM polarization and the electromagnetic field distribution is displayed in \cref{fig1}(b) and \cref{fig1}(c) for TE and TM modes.

\begin{figure}[h]
\centering
\includegraphics[width=1.0\linewidth]{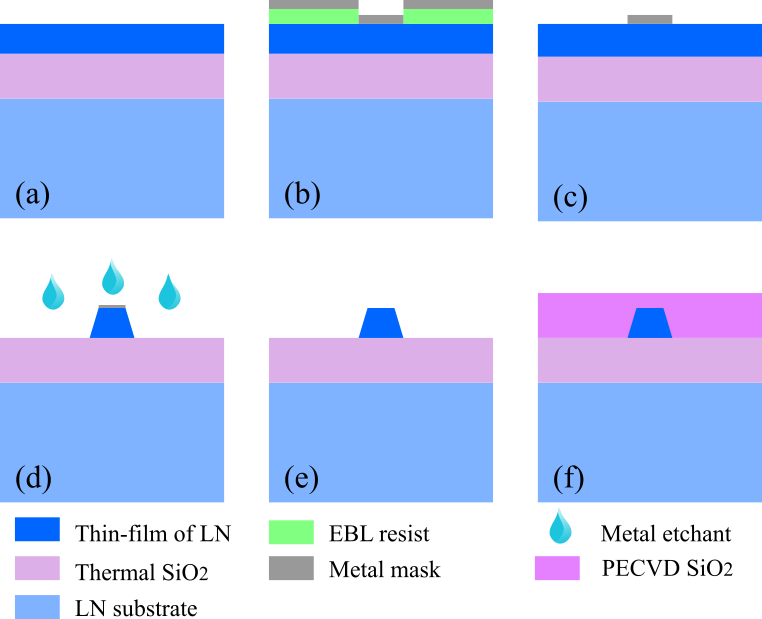}
\caption{Fabrication process of LNOI waveguides: (a) a cross-section of the initial LNOI substrate used for waveguide fabrication; (b) a positive resist is patterned by electron-beam lithography and followed by a metal film layer deposition using an electron beam evaporator; (c) hard metal mask is formed via lift off technique; (d) RIE to transfer the pattern into the Lithium Niobate followed by the metal mask removal via wet etching; (e) etched ridge waveguide without metal mask; (f) PECVD deposition of SiO$_2$ protective layer over fabricated structure.}
\label{fig2}
\end{figure}

The fabrication process implemented to realize the ultra-low loss photonic circuits is shown in \cref{fig2}. The raw sample is a \SI{500}{nm} thick z-cut LN film on top of \SI{2}{\micro\meter} thick buried SiO$_2$ layer supported by a single-crystal z-cut LN substrate; the LNOI wafer was fabricated by Nanoln using the smart-cut technique. We use an optimized lift-off technique to define the circuit pattern and then transfer it to the substrate using RIE. In the first step of the fabrication process, shown in \cref{fig2}, the positive resist is patterned by electron-beam lithography (EBL) to define the mask; the width of patterned structures is \SI{1}{\micro\metre}. The next step consists of metal (Cr) film deposition using an electron beam evaporator, and then lift-off is performed in either acetone or an NMP-based solution. The thickness of the metal layer is optimized to the selectivity of the etching process and depends on the desired etch depth. The chip is then etched in a mix of fluorine and argon plasma allowing smooth and near vertical sidewalls due to the combination of plasma chemical etching and physical sputtering. The metal mask is then removed via wet etching and the final structure cladded with SiO$_2$. The top cladding minimizes the scattering loss induced by the sidewall roughness, facilitates further packaging processes (dicing and CMP polishing) of the final device, and protects the fragile structure from damage.

\section{Results}

\begin{figure}[t]
\centering
\includegraphics[width=1.0\linewidth]{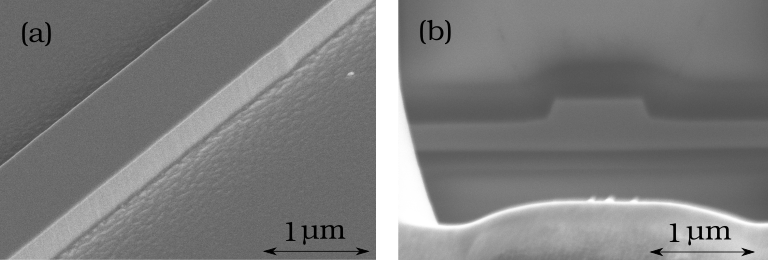}
\caption{Scanning electron microscope images: (a) typical sidewall roughness achieved with optimized fabrication process; (b) cross-section of an optical component discussed in this article.}
\label{fig3}
\end{figure}

The fabricated structures were investigated using optical microscope, surface profilometer, scanning electron microscope (SEM), atomic force microscopy (AFM) and focused ion beam (FIB). The low roughness resulting by our fabrication process is shown in \cref{fig3}(a), AFM was used to confirm the $<$\SI{2}{\nano\meter} RMS. The ridge waveguide cross-section, obtained by focused ion beam milling, shows a sidewall angle of $\sim$\SI{75}{\degree} and an etch depth of \SI{260}{nm}, approximately half the film thickness \cref{fig3}(b).

\begin{figure}[t]
\centering
\includegraphics[width=1.0\linewidth]{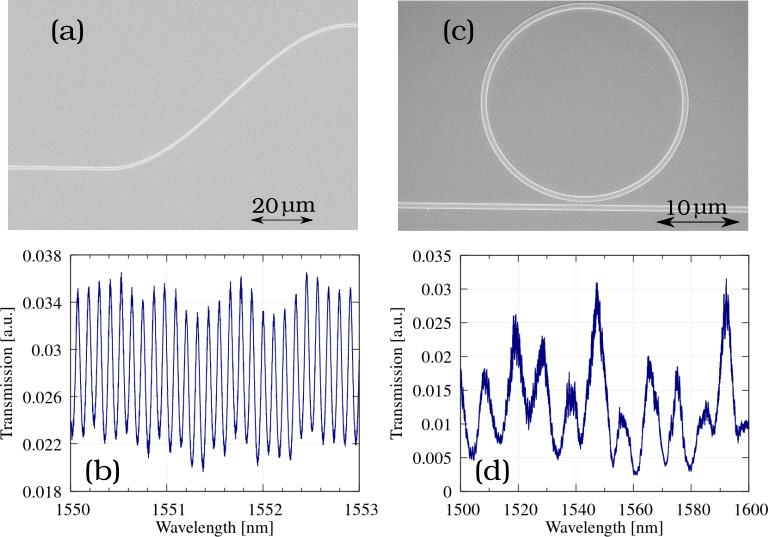}
\caption{(a)SEM image of LNOI s-bend waveguide. (b) Transmission spectrum of s-bend LNOI used for calculating the propagation loss of the TE mode. (c) Optical ring resonator with \SI{15}{\micro\meter} radius and \SI{300}{\nano\meter} gap between ring and bus waveguide used in this experiment. (d) Transmission spectrum of the ring resonator used to determine the effective index of the TE mode.}
\label{fig4}
\end{figure}

The optical loss characterization of a \SI{50}{\micro\meter} s-bend waveguide (shown in \cref{fig4}(a)) is performed using the Fabry-Perot (FP) loss measurement technique \cite{Regener1985}. The bend is used to reduce direct laser light coupling between input and output fibres. Laser light at \SI{1550}{\nano\meter} wavelength is coupled into and out of the polished facets of the waveguide using polarization maintaining (PM) lensed fibers with a mode field diameter of \SI{2}{\micro\meter}. The total input and output coupling and propagation loss is \SI{15}{dB} for a \SI{5}{mm} long chip. The optical transmission spectrum is shown in \cref{fig4}(b). 
The accurate estimation of the waveguide's propagation  loss via the FP technique requires the effective index of the waveguide, which can be found by the free spectral range (FSR) of a ring resonator of known dimensions. For this purpose, we designed and characterized a microring resonator, shown in \cref{fig4}(c), and the optical transmission spectrum is reported in \cref{fig4}(d). The low quality factor of the ring reported here is due to the significant bending loss resulting from the small radius \SI{15}{um} of the ring. High quality factor ring resonators can be easily designed and fabricated by increasing the bend radius.

The effective index $n_\mathrm{eff} = \lambda^2/(L_c\cdot\mathrm{FSR})$, is obtained using the measured FSR = \SI{10}{\nano\meter} (\SI{9.5}{\nano\meter}) for TE (TM), ring radius R = \SI{15}{\micro\meter} and $L_c = 2\pi R + C$, where the coupling length C = \SI{20}{\micro\meter}. The result is $n_\mathrm{eff}$ = 2.00 (2.05) for TE (TM) polarization, which is in good agreement with our FDTD mode solver simulation. 
The propagation loss is calculate to be $\sim0.4\pm0.02\ (0.9\pm0.06)$\,dB/cm for TE (TM) polarization.
The error is calculated from the standard deviation of the fringe's noise.

\section{Discussion}
The transmission measurements show that the optical losses are dominated by the mode mismatch between fiber and waveguide. To improve the coupling efficiency, inverse tapers or grating couplers can be used---the near-vertical angle of our waveguides enables the fabrication of these photonic components. 
Our sidewall angle of $\sim$\SI{75}{\degree} in LNOI waveguides is a significant improvement over previously reported $\sim$\SI{40}{\degree} \cite{Wang:2016uy, Lin:2017} and close to commercially available waveguides in other platforms. Furthermore, to our knowledge, this is the first single mode waveguide reported in z-cut LNOI fabricated via RIE, and the measured sidewall roughness of $<$\SI{2}{nm} RMS is the lowest reported to date in this material. Both sidewall angle and roughness can be further improved by increasing the ratio of chemical etching over physical sputtering.

While LNOI photonic components demonstrated here exhibit ultra-low optical propagation loss, the relatively shallow etching depth results in weak mode confinement, which sets a limit to the minimum bend radius achievable, hence limiting the circuit complexity. 
By increasing the etch depth, we expect to achieve high component density while preserving low propagation loss, which is comparable to AlN and SiN\cite{Tang:2012}. 

\section{Conclusion}
\noindent We have reported the fabrication of ultra-low loss single mode waveguides at telecom wavelength, lower than commercially available Si and SiN photonic platforms. The key advantage of using LNOI over Si and SiN is the second-order nonlinearity (not present in these materials due to the centrosymmetry of their crystalline structures) which enables the implementation of frequency converters, ultra-fast and lossless switches, as well as single photon sources for quantum technology.
Future work will focus on the optimization of the process on x-cut LNOI, which can simplify the poling process, and MgO:LNOI, which has 170x higher optical damage threshold than undoped LN, and can support high power wavelength conversion applications. 

\section*{Funding}
Australian Research Council Centre for Quantum Computation and Communication Technology CE170100012; Australian Research Council Discovery Early Career Researcher Award, Project No. DE140101700; RMIT University Vice-Chancellors Senior Research Fellowship.

\section*{Acknowledgements}
The authors acknowledge Andreas Boes for technical support. This work was performed in part at the Melbourne Centre for Nanofabrication (MCN) in the Victorian Node of the Australian National Fabrication Facility (ANFF) and the Nanolab at Swinburne University of Technology. The authors acknowledge the facilities, and the scientific and technical assistance, of the Australian Microscopy \& Microanalysis Research Facility at RMIT University.

%

\end{document}